\documentclass[submission,copyright,creativecommons,noderivs]{eptcs}
\providecommand{\event}{MBT 2012} 
\usepackage{breakurl}             

\usepackage{graphicx}
\usepackage{algorithm}
\usepackage{algorithmic}
\usepackage{listings}
\usepackage{amssymb}
\usepackage{amsmath}
\usepackage{units} 
\usepackage{amsthm} 

\theoremstyle{theorem}
\newtheorem{theorem}{Theorem}[section]
\theoremstyle{definition}
\newtheorem{example}{Example}[section]
\theoremstyle{theorem}
\newtheorem{definition}{Definition}[section]

\title{Towards Symbolic Model-Based Mutation Testing: Combining Reachability
and Refinement Checking}

\author{Bernhard K. Aichernig \qquad\qquad Elisabeth J\"obstl
\institute{Institute for Software Technology\\
Graz University of Technology\\
Graz, Austria}
\email{aichernig@ist.tugraz.at \qquad\qquad joebstl@ist.tugraz.at}
}
\def\titlerunning{Towards Symbolic Model-Based Mutation Testing}
\def\authorrunning{B.K. Aichernig \& E. J\"obstl}

\def\DEF{\ \ =_{\text{\tiny\it\!df}}\ \ }

\def\=>{\ \Rightarrow\ }
\def\n=>{\ \not\Rightarrow\ }
 
\def\:={\ :=\ }

\def\T0{T_{\!\text{\tiny$\varnothing$}}}

\def\REF{\ \sqsubseteq\ }
\def\Ref{\sqsubseteq}

\def\<{\langle}
\def\>{\rangle}

\newcommand\codesize{\footnotesize \def\@listi{\leftmargin\leftmargini
               \parsep 0\p@ \@plus1\p@ \@minus\p@
               \topsep 8\p@ \@plus2\p@ \@minus4\p@
               \itemsep0\p@}%
    \belowdisplayskip \abovedisplayskip
}
\newcommand{\code}[1]{\texttt{\codesize #1}}
\newcommand{\kw}[1]{\textbf{\code{#1}}} 
\newcommand{\sep}{{~~|~~}} 
\newcommand{\many}[1]{\overline{#1}} 
\newcommand{\utpconcat}[2]{#1\ensuremath{{~\widehat{\ }}}~#2}

\begin{document}
\maketitle

\begin{abstract}
  Model-based mutation testing uses altered test models to derive test
  cases that are able to reveal whether a modelled fault has been
  implemented. This requires conformance checking between the original
  and the mutated model. This paper presents an approach for symbolic
  conformance checking of action systems, which are well-suited to
  specify reactive systems. We also consider non-determinism in our
  models. Hence, we do not check for equivalence, but for refinement.
  We encode the transition relation as well as the conformance
  relation as a constraint satisfaction problem and use a constraint
  solver in our reachability and refinement checking
  algorithms. Explicit conformance checking techniques often face
  state space explosion. First experimental evaluations show that our
  approach has potential to outperform explicit conformance checkers.
\end{abstract}

\section{Introduction}
In most cases, full verification of a piece of software is not
feasible. Possible reasons are the increasing complexity of software
systems, the lack of highly-educated staff or monetary restrictions.
In order to ensure quality and validate system requirements, testing
is a viable alternative if it is systematic and automated. Model-based
testing fulfills these criteria. The test engineer creates a formal
model that describes the expected behaviour of the system under test
(SUT). Test cases are then (automatically) derived from this test model by
applying different algorithms and test specifications.

One big question is where to get the test specifications from. Our
approach is fault-centred, i.e., mutation-based. Classical mutation
testing is a method to assess and increase the quality of an existing
test suite. The source code of the original program is syntactically
altered by applying patterns of typical programming errors, so-called
\emph{mutation operators}~\cite{DeMillo78, Hamlet77}. The test cases
are then executed on the generated \emph{mutants}. If not at least one
test case is able to kill a mutant, the test suite has to be
improved. Mutation testing relies on two assumptions that have been
empirically confirmed: (1)~The \emph{competent programmer hypothesis}
states that programmers are skilled and do not completely wrong. It
assumes that they only make small mistakes. (2)~The \emph{coupling
  effect} states that test cases which are able to detect simple
faults (like faults introduced by mutations) are also able to reveal
more complex errors.

We employ the mutation concept on the test model instead of the source
code and generate test cases that are able to kill the mutated models
(\emph{model-based mutation testing}). The generated tests are then
run on the SUT and will detect whether a modelled fault has been
implemented. So far, much more effort has been spent on the definition of
mutation operators and classical mutation testing and not so much work
has been done on test case generation from mutations~\cite{Jia11}.

What we have not mentioned so far: It is possible that a mutant does
not show any different behaviour from the original program, although
it has been syntactically changed. In this case, the mutant is
equivalent to the original and no test case exists that can
distinguish the two programs. In general, it is not decidable whether
two programs are equivalent. Hence, mutation testing and its wider
application are constrained by the \emph{equivalent mutants
  problem}~\cite{Jia11}. For test case generation, we also have to
tackle this problem. Only if the original and the mutated model are
not equivalent, we can generate a distinguishing test case. In our
case, we do not check for total equivalence, but for refinement. The
models we use are \emph{action systems}, which were originally
introduced by Back~\cite{BackK1983Decentralization}. Action systems
are well-suited for modelling reactive systems and allow
non-determinism.

Within the European project
MOGENTES~\footnote{\url{http://www.mogentes.eu}}, our group already
developed a test case generation tool named \emph{Ulysses}. It is
basically an \emph{ioco} checker for action systems and performs an
explicit forward search of the state spaces. \emph{ioco} is the input-output
conformance relation by Tretmans~\cite{ioco1996}. Ulysses does not only
work for discrete systems, but also supports hybrid action systems via
qualitative reasoning techniques~\cite{qsic2010}.  Experiments have
shown that the performance of explicit enumeration of the state space
involves high memory consumption and runtimes when being applied on
complex models. In this paper, we present an alternative approach to
determine (non-)refinement between two action systems.

As already shown in~\cite{aichernig_qsic05, mbt09}, constraint
satisfaction problems can be used to encode conformance relations and
generate test cases. Each of this works dealt with transformational
systems, i.e., systems that are started and take some input, process
the input by doing some computations and then return an output and
stop again. As already mentioned, action systems are well-suited to
model reactive systems, i.e., systems that are continuously
interacting with their environment. This kind of systems bring up a
new aspect: reachability. Hence, the main contribution of
this paper is a symbolic approach for refinement checking of reactive
systems via constraint solving techniques that avoids state space
explosion. We use the predicative semantics of action systems to
encode (1) the transition relation and (2) the conformance relation as
a constraint satisfaction problem. The constraint system representing
the transition relation is used for a reachability analysis like it is
known from model checking. For each reached state, we test whether it
fulfills the constraint system that represents the conformance
relation, which is refinement.

The rest of this paper is structured as follows. The next section
presents our running example, a car alarm
system. Section~\ref{sec:prelim} gives an overview of the syntax and
semantics of action systems and introduces the conformance relation we
use. Section~\ref{sec:searchUnsafe} explains our approach for finding
differences between two action systems. Afterwards,
Section~\ref{sec:results} presents some experimental data on the
application of our implementation on the car alarm
system. Subsequently, Section~\ref{sec:futureWork} deals with
restrictions and mentions some of our plans for future work. Finally,
Section~\ref{sec:concl} discusses related work and concludes the paper.

\section{Running Example}
\label{sec:cas}

In order to demonstrate the basic concepts of our approach, we use a
simplified version of a car alarm system (CAS). The example is taken
from Ford's automotive demonstrator within the 
MOGENTES project. The
following requirements were specified and served as the basis for our
model:
\begin{description}
\item[R1 - Arming.] The system is armed 20 seconds
after the vehicle is locked and the bonnet, luggage compartment, and
all doors are closed.

\item[R2 - Alarm.] The alarm sounds for 30 seconds if
an unauthorized person opens the door, the luggage compartment, or the
bonnet. The hazard flasher lights will flash for five minutes.

\item[R3 - Deactivation.] The anti-theft alarm system
can be deactivated at any time, even when the alarm is sounding, by
unlocking the vehicle from outside.
\end{description}

\begin{figure}
  \begin{center}
    \includegraphics[width=0.7\linewidth,keepaspectratio]{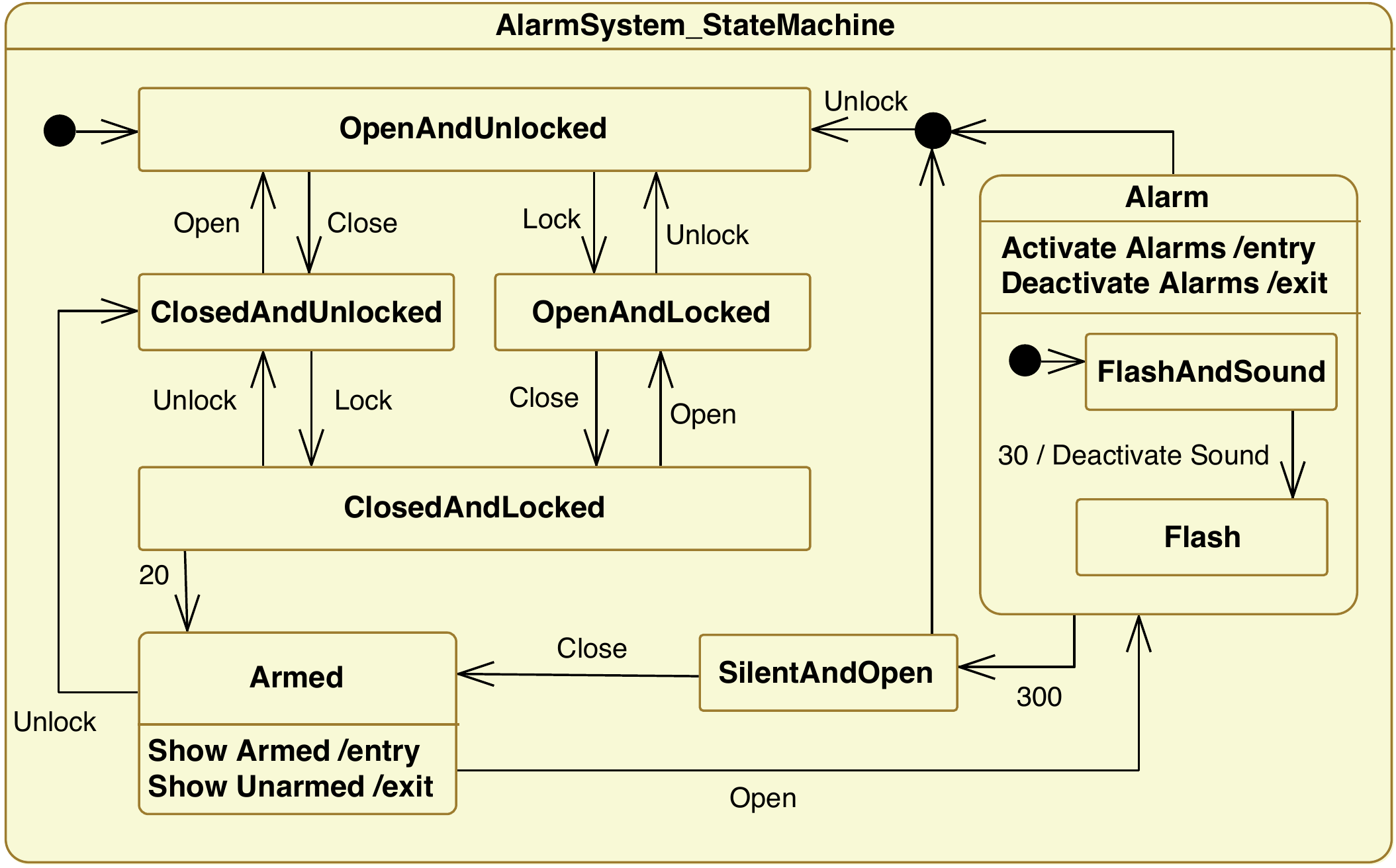}
    \caption{UML state machine of the car alarm system}
    \label{fig:CasStateMachine}
  \end{center}
\end{figure}

Figure~\ref{fig:CasStateMachine} shows a UML state machine
of our CAS. From the state \emph{OpenAndUnlocked} one can traverse to
\emph{ClosedAndLocked} by closing all doors and locking the
car.
As specified in requirement~R1, the alarm system is armed after 20
seconds in \emph{ClosedAndLocked}. Upon entry of the \emph{Armed}
state, the model calls the method \emph{AlarmArmed.SetOn}. Upon
leaving the state, which can be done by either unlocking the car or
opening a door, \emph{AlarmArmed.SetOff} is called. Similarly, when
entering the \emph{Alarm} state, the optical and acoustic alarms are
enabled. When leaving the alarm state, either via a timeout or via
unlocking the car, both acoustic and optical alarm are turned
off. Note that the order of these two events is not specified, neither
for enabling nor for disabling the alarms. Hence
the system is not deterministic. When leaving the alarm state
after a timeout (cf. requirement~R2) the system returns to an armed
state only in case it receives a close signal. Turning off the
acoustic alarm after 30 seconds, as specified in requirement~R2, is
reflected in the time-triggered transition leading to the \emph{Flash}
sub-state of the \emph{Alarm} state.

\section{Preliminaries}
\label{sec:prelim}
\subsection{Action Systems}
\label{sec:as}
Action systems \cite{BackK1983Decentralization} are a kind of
guarded-command language for modelling concurrent reactive
systems. They have a formal semantics with refinement laws and are
compositional \cite{BackSere1991RefinementOfActionSystems}.  Many
extensions exist, but the main idea is that a system state is updated
by guarded actions that may be enabled or not. If no action is
enabled, the action system terminates. If several actions are enabled,
one is chosen non-deterministically. Hence, concurrency is modelled in
an interleaving semantics. The formal method $B$ has recently adopted the
action-system style in the form of \emph{Event-B} \cite{Abrial10}.
\begin{example}
  Our action systems are written in Prolog
  syntax. Listing~\ref{lst:cas} shows code snippets from the action
  system model of the CAS as described in Section~\ref{sec:cas}. The
  first two lines contain user-defined types. All types are basically
  integers, but their ranges can be restricted. In Line~1, a type with
  name \emph{enum\_State} is defined. Its domain begins with 0 and
  ends with 7. Line~4 declares a variable with name \emph{aState}
  which is of type \emph{enum\_State}. Line~6 defines the list of
  variables that make up the state of the action system. The
  \emph{init} predicate in Line~8 defines the initial values for the
  state. At Line~10, the actual action system begins. It consists of
  an \emph{actions} block (Lines 11 to 30) and an \emph{do-od} block
  (Lines 31 to 35).

  The \emph{actions} block defines named actions. Each action consists
  of a name, a guard and a body ($\mathit{name::guard=>body}$) (cf.
  Lines~23 to 28). Actions may also have parameters, like action \emph{after}
  in Line~12. The operator~$[\,]$ denotes non-deterministic choice. We
  use it in our example together with guards to distinguish between
  different cases in which an action may fire. Consider for example
  Lines 14 and 15. The action \emph{after(20)} may fire if the action
  system is in a state where variable $aState$ equals $3$, which
  corresponds to state ``ClosedAndLocked'' in the CAS state chart
  (Figure~\ref{fig:CasStateMachine}). The action system then assigns
  variable \emph{aState} value~2 and variable
  \emph{fromClosedAndLocked OR fromSilentAndOpen} value~1, which
  corresponds to the state ``Armed'' in the state
  chart. The do-od block connects previously defined actions via
  non-deterministic choice. Basically, the execution of an action
  system is a continuous iteration over the do-od block. Here, there
  is always at least one action enabled. Hence, the car alarm system never
  terminates, but continuously waits for stimuli. 

\lstset{language=Prolog,
basicstyle=\footnotesize, 
numbers=left, 
numberstyle=\footnotesize, 
morekeywords={type, var, state_def, init, as, actions, dood}}
\begin{lstlisting}[mathescape,float=t, caption={Code snippet from the action
    system model for the car alarm system},label=lst:cas]
type(enum_State, X) :- X in 0..7.
type(int, X) :- X in 0..270.
...
var([aState], enum_State).
...
state_def([aState, fromAlarm, fromArmed, ..., flashOn, soundOn]).

init([6, 0, 0, 0, 0, 0]).

as :- 
    actions (
    'after'(Wait_time)::(true) =>
    (
       ((Wait_time #= 20 $\verb+#/\+$ aState #= 3) =>
            (aState := 2; fromClosedAndLocked_OR_fromSilentAndOpen := 1))
      []
       ((Wait_time #= 30 $\verb+#/\+$ aState #= 1 $\verb+#/\+$ fromArmed #= 4) => 
            (aState := 0; fromAlarm := 4; fromArmed := 0))
      []
       ((Wait_time #= 270 $\verb+#/\+$ aState #= 0 $\verb+#/\+$ fromAlarm #= 2) => 
            (aState := 7; fromAlarm := 1; fromArmed := 0))
    ),    
    'Lock'::(true) => 
    (
       ((aState #= 6 $\verb+#/\+$ fromAlarm #= 0) => (aState := 5))
      []
       ((aState #= 4 $\verb+#/\+$ fromArmed #\= 1) => (aState := 3; fromArmed := 0))
    ),
...
    ),
  dood (
     'Lock'
  [] [X:int]:'after'(X)
  [] ...
  ).
\end{lstlisting}
\end{example}

\begin{figure}[t] 
\footnotesize
\renewcommand{\arraystretch}{1.2} 
\[ \begin{array}{l@{\hspace{1mm}}l@{\!}l@{\hspace{10mm}}l@{\hspace{1mm}}l@{\!}l} 
M &::=~& D ~\kw{as}~\mbox{:--}~\kw{actions}(\many{A}),~\kw{dood}(P).& 
P &::=~& E \sep E\;[]\;P\\
D &::=& \many{\kw{type}(t,
  X)~\mbox{:--}~X~\kw{in}~n_1..n_2.}~~\many{\kw{var}([\many{v}],t).}~~\kw{state\_def}([\many{v}]).~\kw{init}([\many{c}]). &
E  &::=&  l \sep [\many{X : t}]l(\many{X})\\
A &::=& L::g => B& 
L &::=&  l \sep l(\many{X})\\
B &::=& v := e  \sep g  => B \sep B ; B \sep B\;[]\; B&
e &::= & v \sep c \sep e + e \sep ...\\
\end{array} \] 
\caption{Syntax of a subset of action systems}\label{syntax}
\end{figure} 

\paragraph{Syntax.} In the literature many versions of Back's original
action-system notation \cite{BackK1983Decentralization} exist. The syntax used in this work is
presented in Figure~\ref{syntax}. Our syntax contains some elements of
Prolog, because the tool is implemented in SICStus Prolog. Here, an
action system model $M$ comprises the basic definitions $D$, a set of
action definitions $\many{A}$ and the do-od block $P$. In the basic
definitions we define the types $t$, declare variables $v$ of type
$t$, define the system state-space as variable vector $\many{v}$ and finally
provide the initial state as vector of constants $\many{c}$. An action
$A$ is a labelled guarded command with label $L$, guard $g$ and body
$B$. Actions may have a list of parameters $\many{X}$.  The body of an
action may assign an expression $e$ to a variable $v$ or it may be
composed of (nested) guarded commands itself. Composition may be
sequential or non-deterministic choice. The do-od block $P$ provides
the event-based view on the action system. Here, the actions are
composed by their action labels $l$. Currently, we only support
non-deterministic choice in the do-od block, but in future sequential
and prioritized composition will be added.
  
\begin{figure}[t] 
\footnotesize
\renewcommand{\arraystretch}{1.2} 
\[ \begin{array}{lll@{\hspace{10mm}}lll} 
l::g => B & =_{\tiny\mbox{\it df}}& g~\wedge~B~\wedge~tr' =
\utpconcat{tr}{[l]} & 
l(\many{X})::g => B & =_{\tiny\mbox{\it df}} &  \exists~\many{X} : g~\wedge~B~\wedge~tr' =
\utpconcat{tr}{[l(\many{X})]} \\
x := e  & =_{\tiny\mbox{\it df}} &  x' = e~\wedge~y' = y~\wedge
... \wedge~z' = z &
g  => B   & =_{\tiny\mbox{\it df}} &  g~\wedge~B\\
B(\many{v},\many{v}') ; B(\many{v},\many{v}')  & =_{\tiny\mbox{\it
    df}} & \exists~\many{v_0} : B(\many{v},\many{v_0})~\wedge~B(\many{v_0},\many{v}' )& 
B\;[]\; B & =_{\tiny\mbox{\it df}} & B~\vee~B
\end{array} \] 
\caption{Predicative semantics of actions}\label{semantics}
\end{figure} 

\paragraph{Semantics.} The formal semantics of action systems is
usually defined in terms of weakest preconditions. However, for our
constraint-based approach, we found a relational predicative
semantics being more suitable. We follow the style of He and Hoare's
Unifying Theories of Programming \cite{HH98}. Figure~\ref{semantics}
presents the formal semantics of the actions of our modelling
language. The state-changes of actions are defined via predicates
relating the pre-state of variables $\many{v}$ and their post-state
$\many{v}'$. Furthermore, the labels form a visible trace of events
$tr$ that is updated to $tr'$ whenever an action runs through. Hence,
a guarded action's transition relation is defined as the conjunction
of its guard $g$, the body of the action $B$ and the adding of the
action label $l$ to the previously observed trace. In case of
parameters $\many{X}$, these are added as local variables to the
predicate. An assignment updates one variable $x$ with the value of an
expression $e$ and leaves the rest unchanged. Sequential composition
is standard: there must exist an intermediate state $\many{v_0}$ that
can be reached from the first body predicate and from which the second
body predicate can lead to its final state. Finally, non-deterministic
choice is defined as disjunction. The semantics of the do-od block is
as follows: while actions are enabled in the current state, one of the
enabled actions is chosen non-deterministically and executed. An
action is enabled in a state if it can run through, i.e. if a
post-state exists such that the semantic predicate can be
satisfied. The action system terminates if no action is enabled.
The labelling of actions is non-standard and has been added in order
to support an event-view for testing.

\subsection{Conformance}
Once the modelling language with a precise semantics is fixed, we can
define what it means that a SUT conforms to a given reference model,
i.e. if the observations of a SUT confirm the theory induced by a
formal model. This relation between a model and the SUT is called the
conformance relation.

In model-based mutation testing, the conformance
relation plays an additional role. It defines if a syntactic change
in a mutant represents an observable fault, i.e. if a mutant is
equivalent or not. However, for non-deterministic models an
equivalence relation is no suitable conformance
relation. An abstract non-deterministic model may do more than its
concrete counterpart. Hence, useful conformance relations are order-relations
rather than equivalence relations, the order going from abstract to
more concrete models. In this work, we have chosen UTP's refinement
relation as a conformance relation. UTP defines refinement via
implication,
 i.e. more concrete implementations $I$ imply more abstract
models $M$.  
\begin{definition}\textbf{(Refinement)}\label{refinement}
\[ M \REF I \DEF  \forall x, x' y, y', \dots \in \alpha\ :\ I \=>
M ~~~~~~~~~~\text{for all $M, I$ with alphabet $\alpha$.} \] 
The alphabet $\alpha$ is the set of variables denoting
observations.
\end{definition}

In \cite{AichernigHe09} we have developed a mutation testing theory based on this
notion of refinement. The key idea is to find test cases whenever a
mutated model $M^M$ does not refine an original model $M^O$, i.e. if $M^O
\not\Ref M^M$. Hence, we are interested in counter-examples to
refinement. From Definition~\ref{refinement}  follows that such
counter-examples exist if and only if implication does not hold:
\[ \exists x, x', y, y',  \dots \in \alpha\ :\ M^M \wedge \neg M^O\]
This formula expresses that there are observations in the mutant $M^M$
that are not allowed by the original model $M^O$. We call a state,
i.e. a valuation of all variables, \emph{unsafe} if such
an observation can be made. 

\begin{definition}\textbf{(Unsafe State)}
A pre-state $u$ is called unsafe if it shows wrong
(not conforming) behaviour in a mutated model $M^M$ with respect to an
original model $M^O$. 
Formally, we have:
\[ u \in \{ s~|~\exists~s' : M^M(s,s') \wedge \neg M^O(s,s')\} \]
\end{definition}
We see that an unsafe state can lead to an incorrect next state.  In
model-based mutation testing, we are interested in generating test
cases that cover such unsafe states. Hence, our fault-based testing
criteria are based on the notion of unsafe states.  How to search for
unsafe states in action systems efficiently is discussed in the next
section.

\section{Searching Unsafe States}
\label{sec:searchUnsafe}

\begin{figure}
  \begin{center}
    \includegraphics[width=1\linewidth,keepaspectratio]{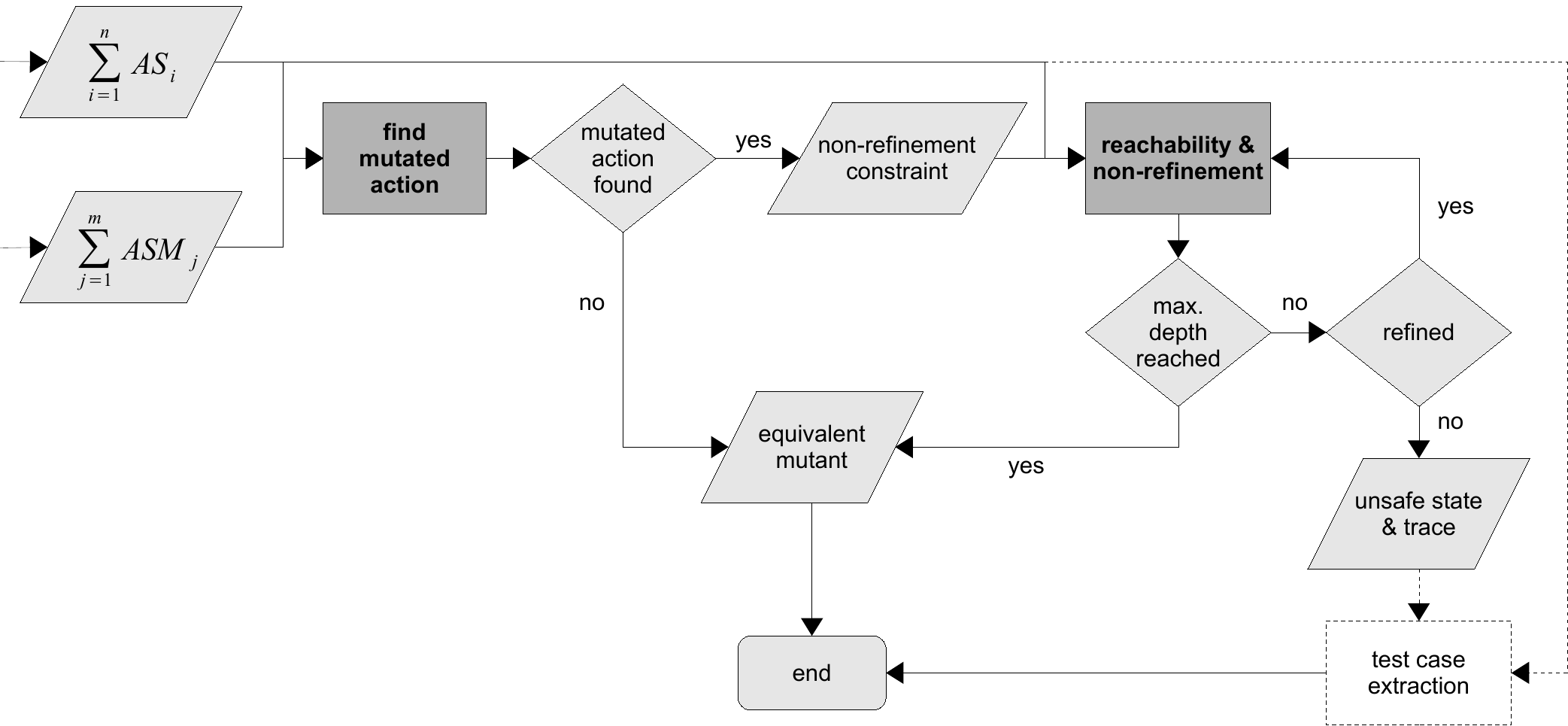}
    \caption{Process for finding an unsafe state}
    \label{fig:process}
  \end{center}
\end{figure}

Figure~\ref{fig:process} gives an overview of our approach to find an
unsafe state. The inputs are the original action system model $AS^O$ and
a mutated version $AS^M$. Each action system consists of a set of
actions $AS^O_i$ and $AS^M_j$ respectively, which are connected via
non-deterministic choice. The first step is a preprocessing activity
to check for refinement quickly. It is depicted on the left-hand side
of Figure~\ref{fig:process} as box \emph{find mutated action}. If
there does not exist an unsafe state at this point, we cannot find
any mutated action that yields non-conformance. Hence, we already know
that the action systems are equivalent. If we find an unsafe state in
this phase, we cannot be sure that it is reachable from the initial
state of the action system. But we know which action has been mutated
and are able to construct a \emph{non-refinement constraint}, which
describes the set of all unsafe states. The next step performs a
reachability analysis and uses the non-refinement constraint to test
each reached state whether it is an unsafe state. In the following, we
give more details.

\subsection{Non-Refinement of Action Systems}
In the previous section, we have introduced non-refinement as a
general criterion for identifying unsafe states. Now, we are going to
concentrate on the special case of action systems.

The observations in our action system language are the event-traces
and the system states before ($\many{v},tr$) and after one execution
$(\many{v}',tr')$ of the do-od block. Then, a mutated action system
$AS^M$ refines its original version $AS^O$ if and only if all
observations possible in the mutant are allowed by the original.
Hence, our notion of refinement is based on both, event traces and
states. However, in an action system not all states are
reachable from the initial state. Therefore, reachability has to be
taken into account.

We reduce the general refinement problem of action systems to a
step-wise simulation problem only considering the execution of the
do-od block from reachable states: 

\begin{definition}\textbf{(Refinement of Action Systems)} 
Let $AS^O$ and $AS^M$ be two action systems with corresponding do-od
blocks  $P^O$ and $P^M$. Furthermore, we assume a function ``reachable''
that returns the set of reachable states for a given trace in an
action system. Then  
\[AS^O \ \REF AS^M \DEF  \forall \many{v},\many{v}',tr,tr' \ :\
((\many{v} \in reachable(AS^O,tr) \wedge P^M) \=>
 P^O)~~~.\] 
\end{definition}
This definition is different to Back's original refinement
definition based on state
traces\cite{BackSere1991RefinementOfActionSystems}.  Here, also the
possible event traces are taken into account. Hence, also the action
labels have to be refined.

Negating this refinement definition and considering the fact that the 
do-od block is a non-deterministic choice of actions $A_i$ leads to the
non-refinement condition for two action systems:
\[
\exists \many{v},\many{v}',tr,tr' \ :\  (\many{v} \in reachable(AS^O,tr) \wedge (A^M_1 \vee \dots \vee A^M_n) \wedge \neg A^O_1 \wedge \dots \wedge
\neg A^O_m)
\]
By applying the distributive law, we bring the disjunction outwards and
obtain a set of constraints for detecting non-refinement.
\begin{theorem}\textbf{(Non-refinement)}\label{nonrefinement} 
A mutated action system $AS^M$ does not refine its original $AS^O$,
iff any action $A^M_i$  of the mutant 
shows trace or state-behaviour that is not possible in the original
action system:
\[AS^O \ \not\Ref AS^M \textbf{~~~iff~~~}  \bigvee^n_{i = 1}~~ \exists
\many{v},\many{v}',tr,tr' \ :\  (\many{v} \in reachable(AS^O,tr)
\wedge A^M_i \wedge \neg A^O_1 \wedge \dots \wedge
\neg A^O_m)\] 
\end{theorem}

In the following, we discuss how this property is applied in our
refinement checking process.

\subsection{Finding a Mutated Action}
\label{sec:find_mutation}

The non-refinement condition presented in Theorem~\ref{nonrefinement} is
a disjunction of constraints of which each deals with one action
$A^M_i$ of the mutated action system $AS^M$.  Hence, it is sufficient
to satisfy one of these sub-constraints in order to find
non-conformance. We use this for our implementation as we perform the
non-refinement check action by action. Here, we first concentrate on
finding a possibly unreachable unsafe state.  Reachability is dealt
with separately (see Section~\ref{sec:reach_unsafe}).

Algorithm~\ref{alg:findMutAct} gives details on the action-wise
non-refinement check, which is depicted on the left-hand side of
Figure~\ref{fig:process} (box \emph{find mutated action}). We
transform the whole do-od block of the original into a constraint
system according to our predicative semantics of action systems
(Line~1). We then translate one action of the mutated action system
into a constraint system (Line~3).  The non-refinement constraint
$\mathit{CS\_nonrefine}$ is the conjunction of the constraint system
representing the mutated action ($CS\_AS^M_i$) and the negated
constraint system representing the original action system ($\neg
CS\_AS^O$, cf. Line~4). Note that sequential composition involves
existential quantification, which becomes universal quantification due
to negation. Existential quantification is implicit in constraint
systems. Universal quantification would lead to quantified constraint
satisfaction problems (QCSPs) that are not supported by common
constraint solvers. Fortunately, we can resolve this problem by a
normal form that requires that non-deterministic choice is always the
outermost operator and not allowed in nested expressions. In this way,
the left-hand side of a sequential composition is always deterministic
and existential quantification can be eliminated. Our car alarm system
example (cf. Listing~\ref{lst:cas}) already satisfies this normal
form. Otherwise, each action system can be automatically rewritten to
this normal form. This has not yet been implemented.

 \begin{algorithm}[t]
\caption{$\mathit{findMutatedAction(AS^O, AS^M): (AS^M_i, CS\_nonrefine)}$}
\label{alg:findMutAct}
\begin{algorithmic}[1]
\STATE $\mathit{CS\_AS^O := trans(AS^O)}$
\FORALL {$\mathit{A^M_i \in AS^M}$}
    \STATE $\mathit{CS\_AS^M_i := trans(A^M_i)}$
    \STATE $\mathit{CS\_nonrefine := CS\_AS^M_i \wedge \neg CS\_AS^O}$
    \IF {$\mathit{sat(CS\_nonrefine)}$}
        \RETURN $\mathit{(A^M_i,
          CS\_nonrefine)}\;\;\;\;\;\;\;$\small{\textit{// mutated action found}}
    \ENDIF
\ENDFOR
\RETURN $\mathit{(nil,\;false)}\;\;\;\;\;\;\;$\small{\textit{// equiv}}
\end{algorithmic}
\end{algorithm}

The non-refinement constraint for the just translated action is then
given to a constraint solver to check whether it is satisfiable by any
$\many{v}, \many{v}', tr, tr'$(Line~5), i.e., whether there exists an
unsafe state $\many{v}$ for $AS^M$ and $AS^O$. If yes, we found the
mutated action and return it together with the according
non-refinement constraint $\mathit{CS\_nonrefine}$. Otherwise, the
next action $A^M_i$ is investigated (loop in Line~2). If no action
leads to a satisfiable non-refinement constraint, then $AS^M$ refines
$AS^O$ (Line~9). Algorithm~\ref{alg:findMutAct} is sound for first
order mutants (one syntactical change per mutant). It aborts after
finding the first action that leads to an unsafe state. Note that we
do not know yet whether an unsafe state is actually reachable. For
higher-order mutants (more than one syntactical change per mutant) it could
happen that our algorithm finds a mutated action for which no unsafe
state is reachable. In this case, it is necessary to go back and search for another
mutated action until an unsafe state is actually reachable or all
actions are processed. 

Identifying the mutated action is important for our performance for
two reasons: (1) Solving the non-refinement constraint
$\mathit{CS\_nonrefine}$ for one action is by far faster than solving
a non-refinement constraint encoding all actions of the mutated action
system at once. Experiments showed that the latter is impractical with
the currently used constraint solver. (2) By knowing which action has
been mutated, we know which non-conformance constraint has to be
fulfilled by an unsafe state. This saves constraint solver calls
during the reachability analysis, which is presented in the following.

\subsection{Reaching an Unsafe State}
\label{sec:reach_unsafe}
Now we know whether there exists any unsafe state. If this is the
case, we also know which action has been mutated and we have
determined a non-refinement constraint that describes the set of all
possible unsafe states. But we do not know yet, whether an unsafe
state is actually reachable from a given initial state.  It is
possible that an unsafe state exists theoretically and has been found
in the previous step, but that no unsafe state is reachable from the
initial state of the system. In this case, the mutated action system
conforms to the original, i.e., the mutant refines the
specification. To find out whether an unsafe state is actually
reachable, we perform a state space exploration of the original action
system $AS$. During this reachability analysis, each encountered state is
examined if it is an unsafe state. This test is realized via a
constraint solver that checks whether the reached state fulfills our
non-refinement constraint (see right-hand side of
Figure~\ref{fig:process}).

\begin{algorithm}[t]
\caption{$\mathit{reachNonRefine(AS^O, CS\_nonrefine, max, init): (unsafe, trace)}$}
\label{alg:reach_nonrefine}
\begin{algorithmic}[1]
\IF {$\exists a,s : \mathit{CS\_nonrefine(init, a, s)}$}
    \RETURN $\mathit{(init, [])}$
\ENDIF
\STATE $\mathit{Visited := \{init\}}$
\STATE $\mathit{ToExplore := enqueue((init, []), [])}$
\WHILE {$\mathit{ToExplore \ne []}$}
    \STATE $\mathit{(s_0, tr\_{s_0}) := head(ToExplore)}$
    \STATE $\mathit{ToExplore := dequeue(ToExplore)}$
    \IF {$\mathit{length(tr\_s_0) < max}$}
        \FORALL{$\mathit{(s_1, a_1) \in succStateAndAction(s_0) : s_1 \not \in Visited}$}
            \STATE $\mathit{tr\_s_1 := add(tr\_s_0, a_1)}$
            \IF {$\mathit{\exists a_2, s_2 : CS\_nonrefine(s_1, a_2, s_2)}$}
                \RETURN $\mathit{(s_1, tr\_s_1)}\;\;\;\;\;\;\;$\small{\textit{// unsafe state}}
            \ENDIF
                \STATE $\mathit{Visited := add(s_1, Visited)}$
                \STATE $\mathit{ToExplore := enqueue((s_1, tr\_s_1), ToExplore)}$
        \ENDFOR
    \ENDIF
\ENDWHILE
\RETURN $\mathit{(nil, [])}\;\;\;\;\;\;\;$\small{\textit{// equiv}}
\end{algorithmic}
\end{algorithm}

The pseudo-code in Algorithm~\ref{alg:reach_nonrefine} gives more
details on our combined reachability and non-refinement check. The
algorithm requires the following inputs: (1) the original action
system $\mathit{AS^O}$, (2) the constraint system
$\mathit{CS\_nonrefine}$ representing the non-refinement constraint
obtained from Algorithm~\ref{alg:findMutAct},
(3) an integer $\mathit{max}$ restricting the search depth, and (4)
the initial state $\mathit{init}$ of the action system $AS^O$. The
algorithm returns a pair consisting of the found unsafe state and the
trace leading there.

At first (Lines 1 to 3), we check whether the initial state is already
an unsafe state. This is, we call the constraint solver with the
non-refinement constraint and set the input state to be the initial
state of $\mathit{AS}$. If the solver finds an action $a$ leading to a
post-state $s$ then we detected non-conformance. We found either a state
that can be reached from $init$ only in the mutant but not in the
original or an action that is enabled at state $init$ only in the
mutant but not in the original. In this case, $init$ is returned as
unsafe state together with the empty trace. Otherwise, we perform a
breadth-first search (Lines 4 to 19) starting at $\mathit{init}$.  The
queue $\mathit{ToExplore}$ holds the states that have been reached so
far and still have to be further explored. It contains pairs consisting of the
state and the shortest trace leading to this state. The set $Visited$ holds
all states that have been reached so far and is maintained to avoid
the re-exploration of states.  To ensure termination, the state space
is only explored up to a user-defined depth $max$ (Line~9).

The function $\mathit{succStateAndAction(s_0)}$ (Line~10) returns the
set of all successors of state $s_0$. Each successor is a pair
consisting of the successor state $s_1$ and the action $a_1$ leading
from $s_0$ to $s_1$.  The successors are calculated via the
predicative semantics of our action systems
(cf. Section~\ref{sec:as}). Thereby, we gain a constraint system
representing the transition relation of our action system. It
describes one iteration of the do-od block. The interesting
variables in the constraint system are the input state variables, the
action variable, and the post-state variables. The input state
variables are set to be equal to the variables in $s_0$. We then use a
constraint solver to set the action variable $a_1$ and the variables
that make up the post-state $s_1$. By calling the constraint solver
multiple times with an extended constraint system (with the added restriction
that the next solution has to be different fromt the previous ones),
we get all transitions that are possible from $s_0$.


Each state $s_1$ that is reached in this way and has not yet been
processed ($\mathit{s_1 \not \in Visited}$) is checked for being an
unsafe state (Line~12). This works analogously to Line~1. If an unsafe
state is found it is returned together with the trace leading there
(Line~13). Otherwise, the state is included in the set of visited
states (Line~15) and enqueued for further exploration (Line~16). If no
unsafe state is found up to depth $max$, the mutant refines the
original action system and we return the pair $(nil, [])$ as a result
(Line~20).

\subsection{Test Case Extraction}
\label{sec:tcg}
We implemented our technique in SICStus
Prolog\footnote{\url{http://www.sics.se/sicstus/}} (version
4.1.2). SICStus comes with an integrated constraint solver
\emph{clpfd} (Constraint Logic Programming over Finite
Domains)~\cite{clpfd}, which we used. Our implementation results
either in the verdict \emph{equiv}, which means that the mutated
action system conforms to the original, or in an unsafe state and a
sequence of actions leading to this state. In the latter case it is
possible to generate a test case. The trace resulting from our
approach is not yet a test case, although it reaches the unsafe
state. We still need to add verdicts (pass, fail, and inconclusive)
where necessary. Additionally, the trace has to be at least one step
longer in order to check that only correct behaviour occurs after the
unsafe state. A test case generated in this way is able to reveal
whether the model mutant has been implemented. This test case
extraction step has not yet been implemented and remains future
work. It is indicated by the dotted parts at the right bottom of
Figure~\ref{fig:process}. For an explicit \emph{ioco} checking
technique, we have suggested different test case extraction strategies
in~\cite{icst2011}.


\section{Empirical Results}
\label{sec:results}
For an empirical evaluation of our prototypical implementation, we
have modelled the car alarm system (CAS) described in
Section~\ref{sec:cas} as an action system. Some code snippets of the
model have already been presented in
Listing~\ref{lst:cas}. Additionally, we have manually created first
order mutants (one mutation per mutant) for the original CAS model. We
applied the following three mutation operators:
\begin{itemize}
\item \emph{guard true}: Setting all possible guards to true resulted
  in 34~mutants.
\item \emph{comparison operator inversion}: The action system
  contains two comparison operators: equality (\texttt{\#=}) and inequality
  (\texttt{\#\textbackslash=}).  Inverting all possible equality operators (resulting in
  inequality) yielded 52~mutants. Substituting inequality by equality
  operators resulted in 4~mutants.
\item \emph{increment integer constant}: Incrementation of all integer
  constants by 1 resulted in 116~mutants. Note that at the upper bound
  of a domain, we took the smallest possible value in order to avoid
  domain violations.
\end{itemize}
From these mutation operators, we obtained a total of 206~mutated
action systems. Additionally, we also included the original action
system as an equivalent mutant. Unfortunately, the currently used
constraint solver was not able to handle 12 of the 207~mutants within
a reasonable amount of time during refinement checking without
reachability (see Section~\ref{sec:find_mutation}). We will try another
constraint solver and see if the performance increases. For now we had
to exclude the 12~mutants from our experiments.

We ran our experiments on a machine with a dual-core processor
(2.8~GHz) and 8~GB RAM with a 64-bit operating
system. Table~\ref{tbl:cas} gives information about the execution
times of our \emph{refinement checker} prototype for the remaining
195~mutations. All values are given in seconds unless otherwise
noted. We conducted our experiments for four different versions of the
CAS: (1)~\textbf{CAS\_1}: the CAS as presented in
Section~\ref{sec:cas} with parameter values 20, 30, and 270 for the
action \emph{after}, (2)~\textbf{CAS\_10}: the CAS with parameter
values multiplied by 10 (200, 300, and 2700), (3)~\textbf{CAS\_100}:
the CAS with parameters multiplied by 100, and (4)~\textbf{CAS\_1000}:
the CAS with parameters multiplied by 1000. These extended parameter
ranges shall test the capabilities of our symbolic approach. 
The column \emph{find
  mutated action} shows that checking whether
there possibly exists an unsafe state and which action has been
mutated (see Section~\ref{sec:find_mutation}) is quite fast. The
reachability and non-refinement check (column \emph{reach \&
  non-refine}, see Section~\ref{sec:reach_unsafe}) needs the bigger
part of the overall execution time (column \emph{total}). The four
versions of the CAS differ only in the parameter values and the
domains for the parameters. Our approach takes almost the same amount
of time for all four versions: approximately $1 \nicefrac{3}{4}$
minutes to process all 195~mutants, on average half a second per
mutant, a minimum time per mutant of 0.03 seconds, and a maximum of
about 3~seconds for one mutant.

\begin{table}
  \centering
  \begin{tabular}{ l l | c c c | c c }
    \textbf{CAS version} & & \multicolumn{3}{c |}{\textbf{refinement checker}} &\multicolumn{2}{c}{\textbf{Ulysses}}\\
                                      &  &  \textbf{find mutated action} & \textbf{reach \& non-refine} & \textbf{total} &\textbf{in/out} & \textbf{out}\\
    \hline
    CAS\_1       & total      & 16    & 90    & 106  & 98    & 65\\
                       & average & 0.08 & 0.46 & 0.54 & 0.50 & 0.34\\
                       & min.      & 0.01 & 0.02 & 0.03 & 0.05 & 0.05\\
                       & max.     & 0.30 & 2.80  & 3.10 & 6.30 & 5.33\\
    \hline
    CAS\_10     & total      & 15    & 86    & 101  & 8.8 h     & 7.9 h\\
                       & average & 0.08 & 0.44 & 0.52 & 2.7 min & 2.4 min\\
                       & min.      & 0.01 & 0.02 & 0.03 & 0.45      & 0.36\\
                       & max.     & 0.27 & 2.80 & 3.07 & 2.6 h     & 2.6 h\\
    \hline
    CAS\_100   & total      & 16    & 90    & 106  & - & -\\
                       & average & 0.08 & 0.46 & 0.54 & - & -\\
                       & min.      & 0.01 & 0.02 & 0.03 & - & -\\
                       & max.     & 0.27 & 2.77 & 3.04 & - & -\\
    \hline
    CAS\_1000 & total      & 15    & 85    & 100  & - & -\\
                       & average & 0.08 & 0.44 & 0.52 & - & -\\
                       & min.      & 0.01 & 0.02 & 0.03 & - & -\\
                       & max.     & 0.27 & 2.69 & 2.96 & - & -\\
  \end{tabular}
  \caption{Execution times for our refinement checking tool and the
    \emph{ioco} checker Ulysses applied on four versions of the car alarm
    system. All values are given in seconds unless otherwise noted.}
  \label{tbl:cas}
\end{table}

To have at least a weak reference point for our performance, we have
also utilized our explicit \emph{ioco} checker
\emph{Ulysses}~\cite{icst2011, qsic2010} to generate tests for the
CAS. We have to admit that this comparison is not totally fair, since
Ulysses works quite differently: First of all, Ulysses uses a
different conformance relation named \emph{ioco} (input-output
conformance for labelled transition systems, see~\cite{ioco1996}). We
ran Ulysses in two settings. First, on the CAS with distinguished
input and output actions. The input actions were Close, Open, Lock,
and Unlock. The remaining actions were classified as outputs. Second,
we classified all actions of the CAS as outputs. This setting is
closer to our notion of conformance, since in refinement we do not 
distinguish between input and output actions.
Nevertheless, the conformance relations are still not
identical. In refinement, we only check that an implementation does
not show unspecified behaviour. Hence, an implementation can always do
less than specified. In \emph{ioco}, abscence of (output) behaviour has to be
explicitly permitted by the specification model.
Another difference between Ulysses and our approach are the
final results. Ulysses generates adaptive test cases, not only a trace
leading to an unsafe state as our tool does
(cf. Section~\ref{sec:tcg}).

Despite these inconsistencies, the comparison with Ulysses still
demonstrates one thing very clearly: the problems with explicit state
space exploration. Ulysses explicitly enumerates all symbolic values
(like parameters in the CAS example). Table~\ref{tbl:cas} also gives
the execution times for Ulysses on the CAS with our two settings: (1)
distinction between inputs and outputs (column \emph{in/out}) and (2)
every action is an output (column \emph{out}).  For the original CAS
version (\textbf{CAS\_1}), Ulysses is faster than our constraint-based
approach, particularly if every action is an output. In this case,
test case generation with Ulysses took only one minute for all
195~mutants. But when it comes to \textbf{CAS\_10} with larger
parameter values (200, 300, and 2700 instead of 20, 30, and 270)
Ulysses runs into massive problems. The execution time drastically
increases to almost 9~hours (in/out) and about 8~hours (out). On
average, each mutant takes 2.7 to 2.4 minutes. One mutant even caused
a runtime of 2.6 hours. We observed a memory usage of up to 6 GB
RAM. We suspect that a significant amount of the execution time is
spent on swapping. For the CAS versions \textbf{CAS\_100} and
\textbf{CAS\_1000}, we did not run Ulysses as the runtimes would be even
higher.

Already for the original CAS (\textbf{CAS\_1}), Ulysses needs 5 to 6
seconds for some mutants that altered the \emph{after} action that has
one parameter: the time to wait with a range from 0 to 270. Our
approach took only 0.1 seconds to find the unsafe state and the
corresponding trace. Hence, Ulysses shows very good performance for
systems with small domains. When it comes to larger ranges of
integers, Ulysses comes to its limits quite soon. In this cases, our
approach represents a viable alternative.

\section{Restrictions and Further Optimizations}
\label{sec:futureWork}
Although our approach shows great promise for solving the problems
with large variable domains, it is far from being perfect. In the
following, we discuss restrictions and possible optimizations of the
overall approach as well as of our current implementation:
More elaborate \emph{conformance relations} are
possible. In \cite{Weiglhofer&10} we presented a predicative semantics for ioco. 
Alternating simulation is also an option.

As already discussed in Section~\ref{sec:tcg}, our approach currently
results in an unsafe state and a trace leading there. The generation
of \emph{adaptive test cases} remains future work.
%
Our action systems are ignorant of \emph{time}. 
In the CAS the waiting time was modelled as a simple parameter.
For more elaborate models with clocks a tick-action modelling the
progress of time is needed. For a full timed-automata model, the
actions could be extended with deadlines similar to~\cite{sifakis}.


One obvious improvement for our \emph{implementation} is the use of more
efficient data structures. Currently, we use lists in most cases as
they are the most common data structure in Prolog. For example, the
set of visited states in Algorithm~\ref{alg:reach_nonrefine} is
currently represented by a list. The use of ordered sets in combination with hash
values would be reasonable. As already mentioned, we implemented our
approach in SICStus Prolog. It comes with a built-in constraint solver
(\emph{clpfd} - Constraint Logic Programming over Finite
Domains~\cite{clpfd}), which we use so far.  Our next steps will
include a comparison with other constraint solvers, e.g.,
Minion\footnote{\url{http://minion.sourceforge.net}}. Additionally, we
already supervise an ongoing diploma thesis on the use of different
SMT solvers like Yices\footnote{\url{http://yices.csl.sri.com/}} or
Z3\footnote{\url{http://research.microsoft.com/en-us/um/redmond/projects/z3/}}.

\section{Conclusion}
\label{sec:concl}
This paper deals with model-based mutation testing. Like in classical
model-based testing, we have a test model describing the expected
behaviour of a system under test. This model is mutated by applying
syntactical changes. We then generate test cases that are able to
reveal whether a software system has implemented the modelled faults.
We have chosen action systems as a formalism for system modelling.  In
this paper, we presented our syntax and a predicative semantics for
action systems. We also explained refinement in the context of action
systems. Most importantly, we have developed and implemented an
approach for refinement checking of action systems as a first step for
test case generation from mutated action systems. Throughout the
whole paper, a car alarm system served as a running example, which was
not only used for illustration but also served as a case study for our
experiments.

We employ constraint satisfaction techniques that have already been
used previously~\cite{aichernig_qsic05, mbt09} to encode conformance
relations and generate test cases. Nevertheless, prior works dealt
with systems that take an input and deliver some output. This paper
deals with refinement checking of reactive systems. The thereby
introduced continuous interaction with the environment brings up a new
aspect: reachability. Hence, the main contribution of this paper is a
symbolic approach for refinement checking of reactive systems via
constraint solving techniques that avoids state space explosion, which
is often a problem with explicit techniques.

Our approach to detect non-refinement in action systems is basically a
combination of reachability and refinement checking. We use the
predicative semantics of action systems to encode (1) the transition
relation and (2) the conformance relation as a constraint satisfaction
problem. During reachability analysis, the constraint system
representing the transition relation is used for finding successor
states. The constraint system encoding refinement enables us to test
each reached state whether it is an unsafe state, i.e., whether this
state is directly followed by observations in the mutant that must not
occur at this state according to the original model.

Experimental results with an action system modelling a car alarm
system have demonstrated the potential of our approach compared to
explicit conformance checking techniques. We conducted experiments
with four different versions of the car alarm system that only differ
in the integer ranges of the parameters. The smallest model deals with
parameters from 0 to 270, the largest model contains integer
parameters from 0 to 270000. Our implementation provides constant
runtime for all four models. For 195 mutated models, we only need
about $1 \nicefrac{3}{4}$ minutes regardless of the parameter
ranges. The explicit conformance checker that we also applied on two
model versions was faster ($1$ to $1 \nicefrac{1}{2}$~minutes)
for the smallest model, but already the next larger model caused an
execution time of about 8~hours.

There is existing literature on model-based mutation testing. One
of the first models to be mutated were predicate-calculus
specifications~\cite{Budd85} and formal
Z~specifications~\cite{Stocks93}. Later on, model checkers were
available to check temporal formulae expressing equivalence between
original and mutated models. In case of non-equivalence, this leads to
counterexamples that serve as test cases~\cite{Ammann98}. This is very
similar to our approach, but in contrast to this state-based
equivalence test, we check for refinement allowing non-deterministic
models. Another conformance relation capable to deal with
non-determinism is the input-output conformance (\emph{ioco}) of
Tretmans~\cite{ioco1996}. The first use of an ioco checker for
mutation testing was on LOTOS specifications~\cite{weiglhofer97}. The
tool \emph{Ulysses} that was already mentioned in
Section~\ref{sec:results} applies ioco checking for mutation-based
test case generation on qualitative action systems~\cite{qsic2010}.
A further conformance relation supporting non-determinism is FDR
(Failures-Divergence Refinement) for the CSP process
algebra~\cite{csp_fdr94}. The corresponding FDR model
checker/refinement checker has been used in~\cite{tcg_fdr08} to set up
a whole testing theory in terms of CSP. This work allows test case
generation via test purposes, but not by model mutation.

Our own past work has shown that typically there is no silver bullet
in automatic test case generation that is able to deal
with every system efficiently~\cite{icst2010}.  As we only used one exemplary
model for evaluating our approach so far, it is too early to say
whether the performance of our approach may be generalized. Future
work will include more experiments with different types of systems to
find this out.

\vspace{-1mm}
\paragraph{Acknowledgment.}
Research herein was funded by the Austrian Research Promotion Agency
(FFG), program line ``Trust in IT Systems'', project number 829583, TRUst via Failed FALsification of Complex
Dependable Systems Using Automated Test Case Generation through Model
Mutation (TRUFAL).

\bibliographystyle{eptcs}
\bibliography{mbt2012}

\end{document}